# The Majorana Neutrinoless Double-Beta Decay Experiment

C. E. Aalseth[1], D. Anderson[1], R. Arthur[1], F. T. Avignone III[2], C. Baktash[3], T. Ball[4], A. S. Barabash[5], R. L. Brodzinski[1], V. Brudanin[6], W. Bugg[7], A. E. Champagne[8,9], Y-D. Chan[10], T. V. Cianciolo[3], J. I. Collar[11], R. W. Creswick[2], P. J. Doe[12], G. Dunham[1], S. Easterday[1], Yu. Efremenko[3,7], V. Egerov[6], H. Ejiri[13], S. R. Elliott[4], J. Ely[1], P. Fallon[10], H. A. Farach[2], R. J. Gaitskell[14], V. Gehman[12], R. Grzywacz[3], R. Hazma[13], H. Hime[4], T. Hossbach[1,2], D. Jordan[1], K. Kazkaz[12], J. Kephart[8,15], G. S. King III[2], O. Kochetov[6], S. Konovalov[5], R. T. Kouzes[1], K. T. Lesko[10], A. O. Macchiavelli[10], H. S. Miley[1], G. B. Mills[4], M. Nomachi[13], J. M. Palms[2], W. K. Pitts[1], A. W. P. Poon[10], D. C. Radford[3], J. H. Reeves[1,2], R. G. H. Robertson[12], R. M. Rohm[8,16], K. Rykaczewski[3], K. Saborov[8,15], Y. Sandukovsky[6], C. Shawley[1], V. Stekhanov[5], W. Tornow[8,16], R. G. van de Water[4], K. Vetter[17], R. A. Warner[1], J. Webb[18], J. F. Wilkerson[12], J. M. Wouters[4], A. R. Young[8,15], and V. Yumatov[4].

*(The Majorana Collaboration)*

[1]The Pacific Northwest National Laboratory, Richland, Washington 9935,USA
[2]Department of Physics and Astronomy, University of South Carolina, Columbia, SC 29208, USA
[3]Oak Ridge National Laboratory, Oak Ridge, Tennessee, 37830, USA
[4]Los Alamos National Laboratory, P-Division, MS H846, Los Alamos, New Mexico 87545, USA
[5]Institute of Theoretical and Experimental Physics, B. Cheremushkinskaya 25, 117259 Moscow, Russia
[6]Joint Institute for Nuclear Research, 141980 Dubna, Russia
[7]Department of Physics and Astronomy, University of Tennessee, Knoxville, Tennessee 37831, USA
[8]Triangle Universities Nuclear Laboratory, Durham, North Carolina 27708, USA
[9]Department of Physics and Astronomy, University of North Carolina, Chapel Hill, NC 27514, USA
[10]Lawrence Berkeley National Laboratory, Berkeley, California 94720, USA
[11]Center for Cosmology, Enrico Fermi Institute, University of Chicago, Chicago, Illinois 60637, USA
[12]Department of Physics, University of Washington, Seattle, Washington 98135, USA
[13]Research Center for Nuclear Physics, Osaka University 567, Japan
[14]Department of Physics, Brown University, Providence, Rhode Island 02912, USA
[15]Department of Physics, North Carolina State University 27695, USA
[16]Department of Physics, Duke University, Durham, North Carolina 27708, USA
[17]Lawrence Livermore National Laboratory, Livermore, California 94551, USA
[18]New Mexico State University, Carlsbad, New Mexico 88220, USA

The proposed Majorana double-beta decay experiment is based on an array of segmented intrinsic Ge detectors with a total mass of 500 kg of Ge isotopically enriched to 86% in $^{76}$Ge. A discussion is given of background reduction by: material selection, detector segmentation, pulse shape analysis, and electro-formation of copper parts and granularity. Predictions of the experimental sensitivity are given. For an experimental running time of 10 years over the construction and operation of Majorana, a sensitivity of $T^{0\nu}_{1/2} \sim 4 \times 10^{27}$ y is predicted. This corresponds to $\langle m_\nu \rangle \sim 0.03 - 0.04$ eV according to recent QRPA and RQRPA matrix element calculations.

*Introduction*

The neutrino mass range of interest favored by the results of neutrino oscillation experiments is now well within the grasp of a well-designed germanium ($^{76}$Ge) zero-neutrino double-beta decay (0νββ-decay) experiment. The observation of this decay mode would be the only known practical experiment to demonstrate that neutrinos are Majorana particles. In the case of Majorana neutrinos, 0νββ-decay is by far the most sensitive way to determine the mass scale of neutrino mass eigenvalues.

Neutrino oscillation data clearly establish that there are three eigenstates that mix and that have mass. The flavor eigenstates, $|\nu_{e,\mu,\tau}\rangle$, are connected to the mass eigenstates, $|\nu_{1,2,3}\rangle$, via a linear transformation:

$$|\nu_l\rangle = \sum_{j=1}^{3} |U^L_{lj}| \, e^{i\delta_j} |\nu_j\rangle, \qquad (1)$$

where $l = e, \mu, \tau$, and the factor $e^{i\delta_j}$ is a CP phase, ±1 for CP conservation.

The decay rate for the 0νββ-decay mode driven by the exchange of a massive Majorana neutrino is expressed as follows:

$$\lambda^{0\nu}_{\beta\beta} = G^{0\nu}(E_0, Z) |\langle m_\nu \rangle|^2 \left| M^{0\nu}_f - (g_A/g_V)^2 M^{0\nu}_{GT} \right|^2, \quad (2)$$

where $G^{0\nu}$ is a factor including phase space and couplings, $|\langle m_\nu \rangle|$ is the Majorana neutrino mass parameter discussed below, $M^{0\nu}_f$ and $M^{0\nu}_{GT}$ are the Fermi and Gamow-Teller nuclear matrix elements respectively, and $g_A$ and $g_V$ are the relative axial vector and vector coupling constants respectively. The mass parameter, $|\langle m_\nu \rangle|$, is the "effective Majorana mass of the electron neutrino." After multiplication by a diagonal matrix of Majorana phases, it is expressed in terms of the first row of the 3×3 matrix of equation (1) as follows:

$$|\langle m_\nu \rangle| \equiv \left| |U^L_{e1}|^2 m_1 + |U^L_{e2}|^2 m_2 e^{i\phi_2} + |U^L_{e3}|^2 m_3 e^{i(\phi_3+\delta)} \right| \quad (3)$$

where $e^{i\phi_{2,3}}$ are the Majorana CP phases (±1 for CP conservation in the lepton sector). These phases do not appear in neutrino oscillation expressions and hence have no effect on the observations of oscillation parameters. The phase angle δ does appear in oscillation experiments. The oscillation experiments have, however, constrained the mixing angles and thereby the $U^L_{lj}$ coefficients in equation (3). Using the best-fit values from the SNO and Super Kamiokande solar neutrino experiments and the CHOZ and Palo Verde reactor neutrino experiments, we arrive at the following expression [Pas02], [Che03], [Bah02], [Avi02]:

$$|\langle m_\nu \rangle| \equiv \left| (0.75^{+0.02}_{-0.04}) m_1 + (0.25^{+0.04}_{-0.02}) m_2 e^{i\phi_2} \right.$$
$$\left. + (<0.026) m_3 e^{i(\phi_3+\delta)} \right|, \quad (4)$$

where the errors were computed from the published confidence level values. The bound on $|U_{e3}|^2$ is at a 95% CL and the errors on the first two coefficients are 1σ.

The results of the solar neutrino and atmospheric neutrino experiments imply the mass square differences $\delta^2_{ij} = |m^2_i - m^2_j|$ but cannot distinguish between two mass patterns (hierarchies): the so called "normal" hierarchy, in which $\delta m^2_{solar} = m^2_2 - m^2_1$ and $m_1 \cong m_2 \ll m_3$, and the "inverted", hierarchy where $\delta m^2_{solar} = m^2_3 - m^2_2$ and $m_3 \cong m_2 \gg m_1$. In both cases $\delta m^2_{AT} \cong m^2_3 - m^2_1$. Considering the values in equation (4), we make the simplifying approximation $|U_{e3}|^2 \ll |U_{e1,2}|^2$ and we set $|U_{e3}|^2 \approx 0$. After a few straightforward algebraic manipulations, and using the central values of equation (4), we can write the following approximate expressions [Avi03]

$$|\langle m_\nu \rangle| \cong m_1 \left| \frac{3}{4} + \frac{1}{4} e^{i\phi_2} \left(1 + \frac{\delta m^2_{solar}}{2 m^2_1}\right) \right|, \quad (5)$$

for the case of "normal" hierarchy, and,

$$|\langle m_\nu \rangle| \cong \sqrt{m^2_1 + \delta m^2_{AT}} \left| \frac{3}{4} e^{i\phi_2} + \frac{1}{4} e^{i\phi_3} \right| \quad (6)$$

in the "inverted" hierarchy case. There is of course no evidence favoring either hierarchy. In Table 1a and Table 1b, we show the predicted central values of $\langle m_\nu \rangle$ as a function of the lightest neutrino mass eigenvalue, $m_1$. These values define the desired target sensitivities of next generation 0νββ-decay experiments. The Majorana $^{76}$Ge experiment is designed to reach deep into the mass range of interest.

**Table 1a. Central values of $|\langle m_\nu \rangle|$ in millielectron volts for the range of interest of $m_1$ ($m_1 < m_2 < m_3$), using the approximate equation (5)[†].**

| Normal Hierarchy ($m_1 \cong m_2 \ll m_3$) | | | |
|---|---|---|---|
| $e^{i\phi_2} = -1$ | | $e^{i\phi_2} = +1$ | |
| $m_1$ | $|\langle m_\nu \rangle|$ | $m_1$ | $|\langle m_\nu \rangle|$ |
| 0 | 2.09 | 0 | 2.09 |
| 20 | 10.0 | 20 | 20.0 |
| 40 | 20.0 | 40 | 40.0 |
| 60 | 30.0 | 60 | 60.0 |
| 80 | 40.0 | 80 | 80.0 |
| 100 | 50.0 | 100 | 100.0 |
| 200 | 100.0 | 200 | 200.0 |
| 400 | 200.0 | 400 | 400.0 |

[†]The value for $m_1=0$ was calculated prior to the expansion.

**Table 1b. Central values of $|\langle m_\nu \rangle|$ in millielectron volts for the range of interest of $m_1$ ($m_1 < m_2 < m_3$), using the approximate equation (6).**

| Inverted Hierarchy ($m_3 \cong m_2 \gg m_1$) | | | |
|---|---|---|---|
| $e^{i\phi_2} = -e^{i\phi_3}$ | | $e^{i\phi_2} = +e^{i\phi_3}$ | |
| $m_1$ | $|\langle m_\nu \rangle|$ | $m_1$ | $|\langle m_\nu \rangle|$ |
| 0 | 22.4 | 0 | 44.7 |
| 20 | 24.5 | 20 | 49.0 |
| 30 | 26.9 | 30 | 53.9 |
| 75 | 43.7 | 75 | 87.3 |
| 100 | 54.8 | 100 | 109.5 |
| 200 | 102.5 | 200 | 204.9 |
| 400 | 201.2 | 400 | 402.5 |

*General Description of the Majorana Experiment*

The proposed Majorana detector is an array of Ge detectors with a total mass of 500 kg of Ge that is isotopically enriched to 86% in $^{76}$Ge. The final configuration is not fixed; however, several have been evaluated with respect to cryogenic performance and background reduction and rejection. This discussion will concentrate on a conventional modular design using ultra-low background cryostat technology developed by the International Germanium Experiment (IGEX). It will also utilize new pulse shape discrimination hardware and software techniques developed by the collaboration and detector segmentation to reduce background.

The most sensitive 0νββ-decay experiments thus far have been the Heidelberg-Moscow [Bau99] and IGEX [Aal02] $^{76}$Ge projects that set lower limits on $T_{1/2}^{0\nu}$ of $1.9 \times 10^{25}$y and $1.6 \times 10^{25}$y respectively. They both utilized Ge enriched to 86% in $^{76}$Ge and operated deep underground. The projection is that the Majorana background will be reduced by a factor of 50 over the early IGEX data prior to pulse shape analysis (from 0.2 counts/keV/kg/y to 0.011 counts/keV/kg/y). This will occur mainly by the decay of the internal background due to cosmogenic neutron spallation reactions that produce $^{56}$Co, $^{58}$Co, $^{60}$Co, $^{65}$Zn, and $^{68}$Ge in the germanium by, limiting the time above ground after crystal growth, careful material selection, and electroforming copper cryostats. One component of the background reduction will arise from the granularity of the detector array. In figure 1, an option for a detector configuration is shown for one module. Each of these modules would have three levels of nineteen detectors in close-packed array. Each detector is 62 mm in diameter and 70 mm long with a mass of ~1.1 kg. In figure 2, an alternative cooling option is shown which clusters all the detectors in a copper vacuum chamber which can then be cooled by immersing the chamber in a vessel of liquid nitrogen or in a jacket of cooled gas.

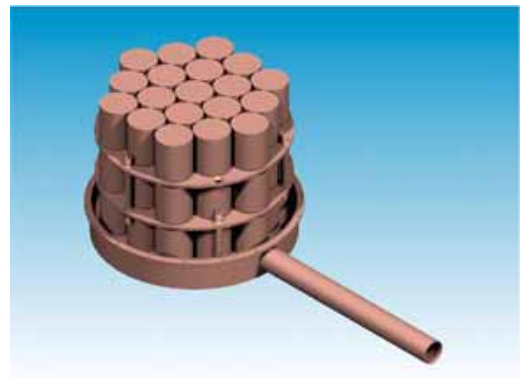

**Figure 1. Possible 57-crystal module. Each crystal is contained in its own copper can.**

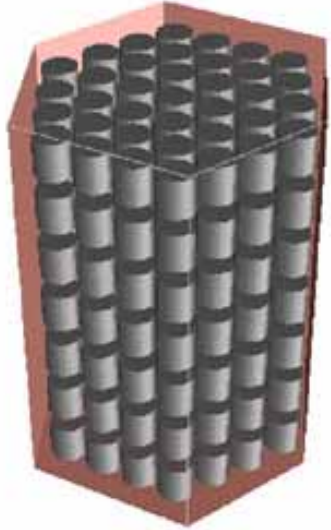

**Figure 2. Alternative cooling scheme.**

*Recent Progress in Ge Detector Technology*

Majorana will not simply be a volume expansion of IGEX. It must have superior background rejection and better electronic stability. The summing of 200 to 250 individual energy spectra can result in serious loss of energy resolution for the overall experiment. In IGEX, instabilities lead to a degradation of 25% in the energy resolution of the 117 mole-years of data. The collaboration has overcome these problems and the technology is now available. First, detectors electronically segmented into 12 individual volumes in a single n-type intrinsic Ge detector are available from two companies: Advanced Measurement Technology (ORTEC) and Canberra Industries. Second, completely digital electronics from XIA (X-ray Instrumentation Associates) has been used by our group to demonstrate unprecedented stability, very low energy thresholds (<1 keV) for a 2-kg Ge detector, and a vast improvement in pulse-shape discrimination.

In the few years since the production of the 2-kg IGEX intrinsic Ge detectors, the new technology evolved in the two industrial companies known to us. Large semi-coaxial n-type detectors have been fitted with a series of azimuthal electrical contacts along their length, and one or more axial contacts in the central hole. A configuration with six-azimuthal-segment by two-axial-segment geometry is shown in Figure 3. After Monte

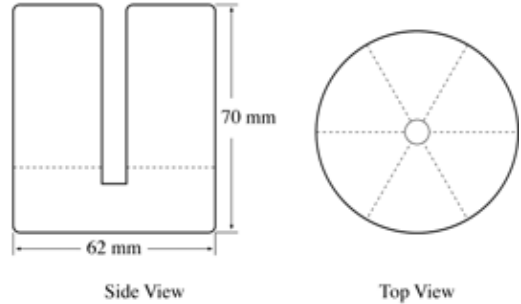

**Figure 3. A Six-by-two twelve-segment detector.**

Carlo studies and discussions with detector manufacturers, several configurations are available that the Majorana collaboration believes strike a good balance between cost, background reduction, and production efficiency. The six-by-two configuration in Figure 3 was used in the Monte Carlo simulations that produced the data shown in Figure 4 for a single detector. The internal $^{60}$Co modeled in the figure is produced by cosmic-ray neutrons during the preparation of the detector. Formation begins after the crystal is pulled. Its elimination by segmentation and pulse shape analysis is crucial.

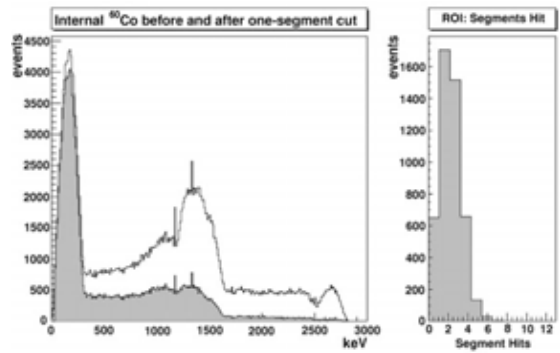

**Figure 4. Monte-Carlo simulation of internal $^{60}$Co background. Left shows a spectrum before and after a one-segment-only cut is applied. Right shows histogram of number-of-segments-hit for events falling in 2.0-2.1 MeV ROI.**

The saga of pulse shape discrimination (PSD) in the IGEX project has been slow and painful, finally culminating in success. Current techniques depend entirely on experimental calibration and do not utilize pulse shape libraries. The ability of these techniques to be easily calibrated for individual detectors makes them practical for large detector arrays.

A major contributor to this success has been the availability of commercial digital spectroscopy hardware. Digitizing a detector preamplifier signal, all subsequent operations on the signal are performed digitally. Programmable digital filters are capable of producing improved energy resolution, long-term stability, and excellent dynamic range. The particular unit used in these studies was the 4-channel Digital Gamma Finder (DGF-4C) unit developed and manufactured by XIA Inc.

The DGF-4C has four independent, 14-bit 40 MHz ADCs. The ADCs are followed by First-in First-out (FIFO) buffers capable of storing 1024 ADC values for a single event. In parallel with each FIFO is a programmable digital filter and trigger logic. The digital filter and trigger logic for each channel is combined into a single field programmable Gate Array (FPGA). Analog input data are continuously digitized and processed at 40 MHz.

The DGF-4C is then a smart filter of incoming pulses. If for example, a signal has a pulse-width incompatible with the usual collection time of 200-300 ns, or is oscillatory (like microphonic noise), the filters can be programmed to reject it. This feature can also be used to allow the very low energy thresholds required in dark matter searches as well as eliminating the broad spectrum of artificial pulses from high-voltage leaks and electromagnetic interference that can even add noise pulses in the region of $0\nu\beta\beta$-decay.

The 14-bit ADCs produce pulse forms that allow the discrimination between single-site interactions in the detector crystal, characteristic of $0\nu\beta\beta$-decay, and the multiple-site interactions characteristic of most gamma-ray background events near 2 MeV. Experimental example pulses are shown in Figure 5. An example single-site event from the 1592 keV double-escape peak of the $^{208}$Tl 2615 keV line is shown as the bottom signal. The top signal is an example multi-site pulse from the full energy peak of the $^{212}$Bi line at 1620 keV.

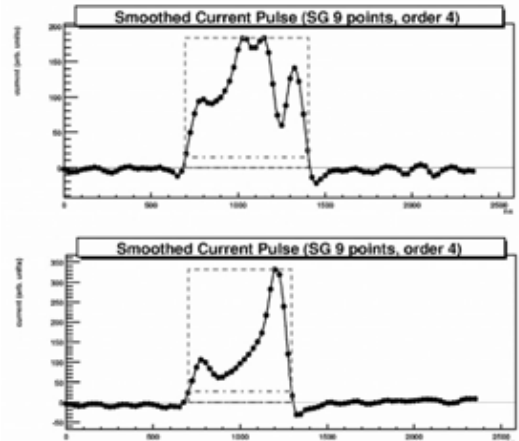

**Figure 5. The top pulse is due to a multiple-location ionization deposit. the bottom pulse is due to a localized deposit.**

*Ultimate Sensitivity of the Majorana Experiment*

To estimate the sensitivity of the Majorana experiment we begin with the published spectrum from an enriched IGEX germanium detector that had been operated under 4000 meters water-equivalent (mwe) shielding from cosmic rays [Bro95]. The components of the background were computed based on the use of validated spallation mechanisms and rates [Avi92]. The computed rate in the region of interest ($R_c = 0.29$ counts/keV/kg/yr) from the spallation isotopes actually exceeded the experimentally measured count rate ($R_e = 0.1$ counts/keV/kg/yr). Therefore, a conservative estimate of 0.2 counts/keV/kg/yr has been taken as an intermediate value. In practice, lower values would be possible by keeping high-energy neutrons away from the raw enriched material and by fabricating the detector underground.

**Table 2. Estimation of sources of activity from early IGEX data and predicted Majorana background.**

| Spallation Isotope | $T_{1/2}$ (d) | Rate from [Bro95] | After Construction | Rate During Experiment | Total in ROI | After PSD Rejection | After Seg Rejection |
|---|---|---|---|---|---|---|---|
| $^{68}$Ge | 270.82 | 0.1562 | 0.03702 | 3.93E-03 | 70.15 | 18.59 | 2.57 |
| $^{56}$Co | 77.27 | 0.0238 | 0.00212 | 6.43E-05 | 1.15 | 0.30 | 0.04 |
| $^{60}$Co | 1925.2 | 0.0177 | 0.01294 | 7.15E-03 | 127.55 | 33.80 | 4.66 |
| $^{58}$Co | 70.82 | 0.0024 | 0.000202 | 5.60E-06 | 0.10 | 0.03 | 0.00 |
| | | cts/keV/kg/y | cts/keV/kg/y | cts/keV/kg/y | Counts | Counts | Counts |
| Total | | 0.2 | 0.0523 | 0.0112 | 198.95 | 52.72 | 7.28 |

It is instructive to scale the count rate of the previous experiment to that of the initial Majorana plan, a 500 kg detector operated for 10 years. We correct that rate to account for the decay of activities that will occur before and during the experiment. Finally, we correct the rate to account for the new technologies that we plan to employ.

The detector used for these sensitivity estimates had been zone refined, so that the $^{60}$Co ($T_{1/2}$ = 5.2 y) inside the crystal, created by cosmic-ray-generated neutrons, was expected to be low. But the detector had been above ground long enough before zone refining to reach equilibrium with respect to $^{68}$Ge ($T_{1/2}$ = 271 d), another important internal contaminant. The first reduction in this background rate comes from decay during the underground array construction period. This has been calculated using a modest rate of production and assuming quarterly shipments of enriched material during the construction period.

Decay during the construction period underground would decrease the $^{68}$Ge by an average factor of 0.24 and an additional factor of 0.11 during the data acquisition of the experiment. Similarly, $^{60}$Co would decay during construction to reduce the count rate to 0.73 of the original rate by the start of the experiment, and during the ten-year data acquisition the average rate during the experiment would be 0.55 times that at the beginning.

Thus, accounting for decay, the average background rate during the experiment would be 0.01 counts/keV/kg/yr. Thus, the effect of pre-deployment decay is effectively a reduction of 94% or a factor of 17.8.

The number of $^{76}$Ge atoms in 500 kg of enriched germanium (86% $^{76}$Ge) is N = 3.41×10$^{27}$. The optimum energy window of δE = 3.568 keV is expected to capture 83.8% of the events in a sharp peak at 2039 keV. If $B = b \cdot \delta E \cdot N \cdot \delta t$ and δt is 10 years, we would expect to observe 199 background counts.

The next step in estimating the sensitivity of the experiment is to apply two new but easily implemented techniques. The first is the pulse shape analysis technique discussed above. This method has been shown to accept $\varepsilon_{PSD}$ = 80.2% of single site pulses (like double-beta decay) and to reject 73.5% of background. The second technique involves the electrical segmentation of the detector crystal to form several smaller segments as discussed earlier. A simplified Monte Carlo analysis, assuming the efficiency $\varepsilon_{PSD}$ is independent of that of segmentation, $\varepsilon_{SEG}$, was carried out only to count the segments with significant energy deposition and reject events with a multiplicity greater than one. This cut accepted $\varepsilon_{SEG}$ = 90.7% of double-beta decay pulses and rejected 86.2% of backgrounds like $^{60}$Co and $^{68}$Ge, which are highly multiple.

Applying the background reduction factors to the simple calculation above, only 7.28 counts of the original 199 counts survive in our 3.568 keV analysis window, a reduction of 96.3% or a factor of 27.3. See Table 2.

The estimated background is ~7.3 events, therefore the sensitivity of the experiment is

~3.8×10²⁷ y at 90% C.L. because we predict that 3.7 events will be the maximum number attributable to 0νββ-decay to a 90% C.L. The computation of the 0νββ half-life must then take into account this number of observable counts, the cut efficiencies, and the fraction of the 0νββ-decay peak found in the analysis window. Thus

$$T_{1/2} = \frac{\ln(2) \cdot N \cdot \Delta t \cdot \varepsilon_{PSD} \cdot \varepsilon_{SEG} \cdot 83.8\%}{3.72} \quad (7)$$
$$= 3.8 \times 10^{27} \, y.$$

A standard relation between the half-life and the effective Majorana mass of the electron neutrino is:

$$\langle m_\nu \rangle = \frac{m_e}{\sqrt{F_N \cdot T_{1/2}}}, \quad (8)$$

where $F_N$ is a nuclear factor computed by various authors. The variety of nuclear calculations gives a range of observable effective Majorana neutrino mass from 0.02 eV to 0.07 eV. Later we present an update of the status of nuclear matrix elements.

Many other formulations of this sensitivity calculation are possible. For instance, it is possible to calculate the expected rate of background due to cosmogenic isotopes in the crystal assuming many different scenarios producing far less initial background. This is a reasonable approach and it would lead to a lower starting background. It is possible, however, to hypothesize away all backgrounds without regard to the effort involved. We chose to start with a known, reproducible starting point so that the result would be credible and attainable. The many details of the technologies involved ranging from lead bricks to multi-dimensional parametric pulse analysis are too lengthy to be described here.

The calculations in this section have covered in some detail the effects of backgrounds on a 5000 kg-y experiment in which the mass is 500 kg and the time is 10 years. A completely different approach would be to consider ways of reducing the time needed to complete the experiment by allowing different total masses of enriched material. In this approach, one might optimize not for lowest cost but for shortest total time to completion, including construction. Many details are not considered in this estimate, such as increased labor costs, increased detector production costs, etc. Figure 6 shows the results of a simple analysis with background rates similar to [Bro95]. Rates of enrichment above 200 kg/y are purely hypothetical, but might be reached by employing more than one Russian enrichment facility.

This simple analysis shows that a significantly reduced schedule is possible with greater investment in enrichment.

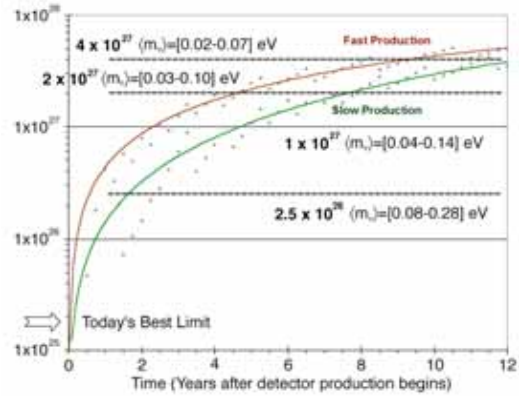

**Figure 6. Sensitivity vs. time of the Majorana Reference Plan using conservative background assumptions. The upper curve assumes a production of germanium of 200 kg/y for a total of 500 kg. Milestones in half-life are shown at 0.25, 1.0, and 4.0×10²⁷ years. Scatter about the trend lines is due to the integer nature of the Poisson distribution.**

*Extraction of the neutrino-mass parameter from double-beta decay half-lives*

The two key goals of 0νββ-decay experiments are: 1) to determine if neutrinos are Majorana particles, and 2) to measure the neutrino-mass-eigenvalues. The latter requires nuclear matrix elements, which must be calculated with specific nuclear models. It is now widely accepted that nuclides that are ββ-decay candidates, $^{76}$Ge, $^{100}$Mo, $^{130}$Te, and $^{136}$Xe for example, are above the nuclear shells where current versions of the nuclear shell model are reliable; however, $^{76}$Ge is probably the best candidate for future shell model calculations of 0νββ-decay matrix elements.

Nevertheless, at present we must rely on schematic models until the development of microscopic models is more advanced.

In 1986, Vogel and Zirnbauer introduced the Quasi-Particle Random Phase Approximation (QRPA) [Vog86]. Since then, there have been many developments and variations, frequently with widely disparate results.

**Table 3. Values of the nuclear structure parameter $F_N \equiv G^{0\nu} | M_f^{0\nu} - (g_A/g_V)^2 M_{GT}^{0\nu} |^2$ calculated with different nuclear models. The effective Majorana mass of the electron neutrino, $\langle m_\nu \rangle$, is given for $T_{1/2}^{0\nu}(^{76}\text{Ge}) = 4 \times 10^{27}$ y.**

| $F_N$ (y$^{-1}$) | $\langle m_\nu \rangle$ eV | Reference |
|---|---|---|
| $1.58 \times 10^{-13}$ | 0.020 | [Hax84] |
| $2.88 \times 10^{-13}$ | 0.015 | [Tom86] |
| $1.12 \times 10^{-13}$ | 0.024 | [Mut89] |
| $1.12 \times 10^{-13}$ | 0.024 | [Sta90] |
| $1.18 \times 10^{-13}$ | 0.024 | [Tom91] |
| $6.97 \times 10^{-14}$ | 0.031 | [Suh92] |
| $7.51 \times 10^{-14}$ | 0.029 | [Suh92] |
| $1.90 \times 10^{-14}$ | 0.059 | [Cau96] |
| $1.42 \times 10^{-14}$ | 0.068 | [Pan96] |
| $7.33 \times 10^{-14}$ | 0.030 | [Pan96] |
| $2.75 \times 10^{-14}$ | 0.049 | [Sim97] |
| $1.33 \times 10^{-13}$ | 0.022 | [Aun98] |
| $8.29 \times 10^{-14}$ | 0.028 | [Fae98] |
| $8.27 \times 10^{-14}$ | 0.028 | [Bar99] |
| $6.19 \times 10^{-14}$ | 0.032 | [Sim99] |
| $2.11 \times 10^{-13}$ | 0.018 | [Sim99] |
| $1.16 \times 10^{-13}$ | 0.024 | [Sto00] |
| $5.22 \times 10^{-14}$ | 0.035 | [Suh00] |
| $2.70 \times 10^{-15}$-$3.2 \times 10^{-15}$ | 0.155-0.143 | [Bob01] |
| $1.80 \times 10^{-14}$-$2.2 \times 10^{-14}$ | 0.060-0.054 | [Bob01] |
| $5.50 \times 10^{-14}$-$6.3 \times 10^{-14}$ | 0.034-0.032 | [Bob01] |
| $1.21 \times 10^{-14}$ | 0.073 | [Sto01a] |
| $1.85 \times 10^{-14}$ | 0.059 | [Sto01a] |
| $3.63 \times 10^{-14}$ | 0.042 | [Sto01a] |
| $6.50 \times 10^{-14}$ | 0.032 | [Sto01a] |
| $7.57 \times 10^{-14}$ | 0.029 | [Sto01b] |

Frequently, bounds on $\langle m_\nu \rangle$ are extracted from experimental limits on $0\nu\beta\beta$-decay half-lives using nuclear matrix elements from all or many available nuclear models. The results can vary by factors of three or more. This is not satisfactory because it does not account for theoretical progress. An example of the variation in extracted values is clearly seen in Table 3.

Until now, conventional wisdom held that knowledge of $2\nu\beta\beta$-decay decay rates would not be useful in determining $0\nu\beta\beta$-decay matrix elements, because the intermediate nuclear states are very different. Recently however, Rodin, Faessler, Simkovic and Vogel showed that in the context of QRPA and Renormalized QRPA (RQRPA) this is not the case [Rod03]. They make a well-documented case that:

**"When the strength of the particle-particle interaction is adjusted so that the $2\nu\beta\beta$-decay rate is correctly reproduced, the resulting $M^{0\nu}$ values become essentially independent on the size of the basis, and on the form of different realistic nucleon-nucleon potentials. Thus, one of the main reasons for variability of the calculated $M^{0\nu}$ within these methods is eliminated".**

Accordingly, one would conclude that accurate measurements of $2\nu\beta\beta$-decay half-lives will have a very meaningful impact on the predictions of $0\nu\beta\beta$-decay matrix elements in the same nuclei. Contrary to previous conventional wisdom, accurate $2\nu\beta\beta$-decay measurements may now be very important in the realm of neutrino physics.

Rodin et al., investigated the effect of the choice of the single-particle (s.p.) space on $M^{0\nu}$, and also used three different realistic nucleon-nucleon interactions utilizing: the Bonn-CD [Mac89], the Argonne [Wir95], and the Nijmegen [Sto94] potentials. The result is that $M^{0\nu}$ varies very little over the 9 different combinations of s.p.-space and interaction.

The effects of neglecting single-particle states further from the Fermi-level were investigated for $^{76}$Ge, $^{100}$Mo, $^{130}$Te, and $^{136}$Xe. In the case of interest here, $^{76}$Ge, the three s.p.-spaces used were: 1) the 9 levels of the oscillator shells N=3 and 4, 2) the addition of the N=2 shell, and finally, 3) the 21 levels from all states in the shells with N=1, 2, 3, 4, and 5. For each change in s.p.-space, the residual interaction must be adjusted by adding a pairing interaction and a particle hole interaction renormalized by an overall strength parameter, $g_{ph}$. The value $g_{ph} \sim 1$ was found to reproduce the giant Gamow-Teller resonance in all cases. Finally, QRPA equations include the effects of a particle-particle interaction, renormalized by an overall strength parameter $g_{pp}$ that in each case was adjusted to reproduce the known $2\nu\beta\beta$-decay rate correctly. Figure 7 clearly shows the unprecedented stability against variations in the model-space and in the realistic nucleon-nucleon interaction used.

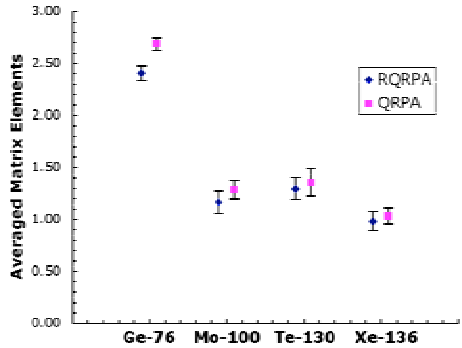

**Figure 7. Results of the nuclear matrix calculations of Rodin et al. [Rod03].**

Finally, we use these results to compute the predicted sensitivity of the Majorana experiment to the effective Majorana mass of the electron neutrino. In the notation of [Rod03], $\langle m_\nu \rangle = [\,|M^{0\nu}|\,(G^{0\nu}\,T^{0\nu}_{1/2})^{1/2}]^{-1}$. They give $|M^{0\nu}| = 2.40 \pm 0.07$ (RQRPA) and $|M^{0\nu}| = 2.68 \pm 0.06$ (QRPA) with $G^{0\nu} = 0.30 \times 10^{-25}$ y$^{-1}\cdot$ eV$^{-2}$. If we choose the round number, $T^{0\nu}_{1/2} = 4 \times 10^{27}$ y for the predicted sensitivity of the Majorana experiment, then the values of the mass parameter corresponding to this half-life are; $\langle m_\nu \rangle = 0.038 \pm 0.007$ eV using RQRPA and $\langle m_\nu \rangle = 0.034 \pm 0.006$ eV with QRPA. A very similar value, $\langle m_\nu \rangle = 0.028 \pm 0.005$ eV, results from using the matrix elements from the recent paper by Civitarese and Suhonen [Civ03].

These values are well within the range of interest tabulated in Table 1, which implies that the Majorana experiment is predicted to reach well into the interesting range of neutrino mass. Should nature have placed the mass below this range, the Majorana array can be expanded and possibly upgraded with newer technology that may emerge, as intrinsic Ge detectors can be re-deployed many times in different configurations.